\documentstyle[sprocl]{article}

\input{epsf}

\def\be{\begin{equation}}
\def\ee{\end{equation}}
\def\bea{\begin{eqnarray}}
\def\eea{\end{eqnarray}}

\def\bbox#1{\hbox{\boldmath{$ {#1} $}}}

\begin{document}

\title {\null\vspace*{-1.7cm}\hfill {\small ORNL-CTP-9805 and
hep-ph/9809497} \\ \vskip 0.8cm Drell-Yan Cross Section and $J/\psi$
Production in High-Energy Nucleus-Nucleus Collisions}


\author{Cheuk-Yin Wong}
\address{Physics Division, Oak Ridge National
Laboratory, Oak Ridge, TN 37831, USA \\E-mail: wongc@ornl.gov} 


\maketitle\abstracts{ We use the Drell-Yan differential distribution
$dN_{DY}^{AB}/dE_T$ in high-energy nucleus-nucleus collisions to
obtain a relation between the transverse energy $E_T$ and the impact
parameter $b$.  Such a relation is then utilized to study the
transverse-energy dependence of $J/\psi$ production in Pb-Pb
collisions, using the $J/\psi$ absorption model presented previously.
The anomalous Pb-Pb suppression data at 158A GeV can be explained if
one assumes the occurrence of a new phase of strong $J/\psi$
absorption when the energy density exceeds 4.2 GeV/fm${}^3$.  The
results are extended to make predictions for $J/\psi$ production at
higher collision energies.  It is found that $J/\psi$ survival
probabilities reach the lowest survival limit when the nucleon-nucleon
center-of-mass energies $\sqrt{s}$ is greater than about 35 GeV. }

\section{Introduction}

Following the initial suggestion by Matsui and Satz \cite{Mat86} that
$J/\psi$ production is suppressed in a quark-gluon plasma, the NA50
experimental observation \cite{Gon96}$^-$\cite{Rom98} of an anomalous
$J/\psi$ absorption in Pb-Pb collisions has led to a flurry of
theoretical activity \cite{Won96qm}$^-$\cite{Arm96}.  The absorption
is anomalous in the systematics of the total normalized yields as a
function of the sum of the radii of the colliding nuclei.  It is also
anomalous in the differential $J/\psi$ survival probability, which
shows a discontinuity as a function of the transverse energy.  A
central question is whether these discontinuities indicate the
occurrence of a new phase of strong $J/\psi$ absorption, as expected
when a quark gluon plasma is produced.

Theoretical studies on $J/\psi$ suppression require the relation
between the transverse energy $E_T$ and the impact parameter $b$ of a
collision.  We shall first show how we can obtain such a relationship
from the Drell-Yan distribution.  We shall use the $J/\psi$ absorption
model presented previously \cite{Won96a} to study $J/\psi$ absorption
in Pb-Pb collisions at $p_{\rm lab}=158$A GeV ($\sqrt{s}=17.3$ GeV).
The results are then utilized to predict $J/\psi$ production at higher
energies where we find interesting trends as the collision energy
increases.

\vspace*{-0.4cm}
\section{  Drell-Yan Cross Section}
\vskip -0.1cm

In NA50 experiments, the transverse energy is measured in coincidence
with the detection of Drell-Yan and $J/\psi$ dimuon pairs.  Based on
the assumption of a monotonic dependence, the relation between $E_T$
and $b$ can be easily obtained from the Drell-Yan differential cross
section $d\sigma_{DY}^{AB}/dE_T$ or the Drell-Yan distribution
$dN_{DY}^{AB}/dE_T$ which is equal to
$(d\sigma_{DY}^{AB}/dE_T)[N_{DY}^{AB}(total)/\sigma_{DY}^{AB}(total)]$.
One relies here on the experimental nuclear mass dependences
\cite{Ald91,Lou96} which indicate that initial- and final-state
interactions have only very small effects on the Drell-Yan cross
section in nuclear collisions.  Neglecting initial- and final-state
interactions, the Drell-Yan cross section element in the collision of
nucleus $A$ with nucleus $B$ is given by
\begin{eqnarray}
d \sigma_{DY}^{AB} = A B ~ T_{AB}(\bbox{b})~ d\bbox{b}~ \sigma_{DY}^{pp},
\end{eqnarray}
where $T_{AB}({\bbox{b}})$ is the thickness function,
\begin{eqnarray}
T_{AB}({\bbox{b}})= \int d\bbox{b}_A T_A (\bbox{b}_A ) T_B
(\bbox{b}-\bbox{b}_A ),
\end{eqnarray}
and $T_A(\bbox{b}_A)$ and $T_B(\bbox{b}_B)$ are the thickness
functions for nucleus $A$ and $B$ respectively \cite{Won94}.  Therefore,
we have
\begin{eqnarray}
\label{eq:dsdet}
{d \sigma_{DY}^{AB} \over dE_T} = A B \sigma_{DY}^{pp}
~T_{AB}(\bbox{b})~ {d\bbox{b} \over d E_T}.
\end{eqnarray}
From these relations, we can get a relation between $E_T$ and $b$
given by
\begin{eqnarray}
\label{eq:etb}
{\Sigma}(E_T) \equiv  {1 \over \sigma_{DY}^{AB}(total)} \int_0^{E_T}
{d\sigma_{DY}^{AB} \over dE_T'} dE_T'
= \int_b^{\infty} T_{AB}(\bbox{b}') d\bbox{b}' \equiv {\Sigma}(b),
\end{eqnarray}
where ${\Sigma}(E_T \rightarrow \infty) = {\Sigma}(b=0) = 1$ and
\begin{eqnarray}
\sigma_{DY}^{AB}(total)=\int_0^{\infty}
{d\sigma_{DY}^{AB} \over dE_T'} dE_T'.
\end{eqnarray}
If the Drell-Yan distribution $dN_{DY}^{AB}/dE_T$ is measured instead,
the function $\Sigma(E_T)$ can be expressed in terms of
$dN_{DY}^{AB}/dE_T$ by
\begin{eqnarray}
{\Sigma}(E_T) =  {1 \over N_{DY}^{AB}(total)} \int_0^{E_T}
{dN_{DY}^{AB} \over dE_T'} dE_T'
\end{eqnarray}
where
\vspace*{-0.5cm}
\begin{eqnarray}
N_{DY}^{AB}(total)=\int_0^{\infty}
{dN_{DY}^{AB} \over dE_T'} dE_T'.
\end{eqnarray}
To obtain $E_T$ as a function of $b$, one can use a graphical method
where one obtains ${\Sigma}(E_T)$ from experimental data and
${\Sigma}(b)$ from $T_{AB}(b)$ which can be calculated from the known
geometry of the colliding nuclei.  One then uses $b$ and $E_T$ as the
abscissa, and plots $\Sigma$ as the ordinate, resulting in the curves
of ${\Sigma(E_T)}$ and ${\Sigma(b)}$.  From Eq.\ (\ref{eq:etb}), the
horizontal line of constant $\Sigma$ then intercepts the two curves at
$b$ and its corresponding $E_T(b)$.  Alternatively, one can
parameterize $E_T(b)$ as
\vspace*{-0.3cm}
\begin{eqnarray}
\label{eq:mean}
E_T(b) = {E_{T0} \over  1 + \exp[ (b-b_0)/a]},
\end{eqnarray}
which gives $db/dE_T$. One then uses Eq.\ (\ref{eq:dsdet}) to
determine $d\sigma_{DY}^{AB}/dE_T$.  The Drell-Yan distribution for
Pb-Pb collisions calculated with $E_{T0}=135$ GeV, $b_0=6.7$ fm, and
$a=3$ fm is shown as the dashed curve in Fig.\ 1.

\null\vskip 5.4cm 
\epsfxsize=250pt
\includegraphics{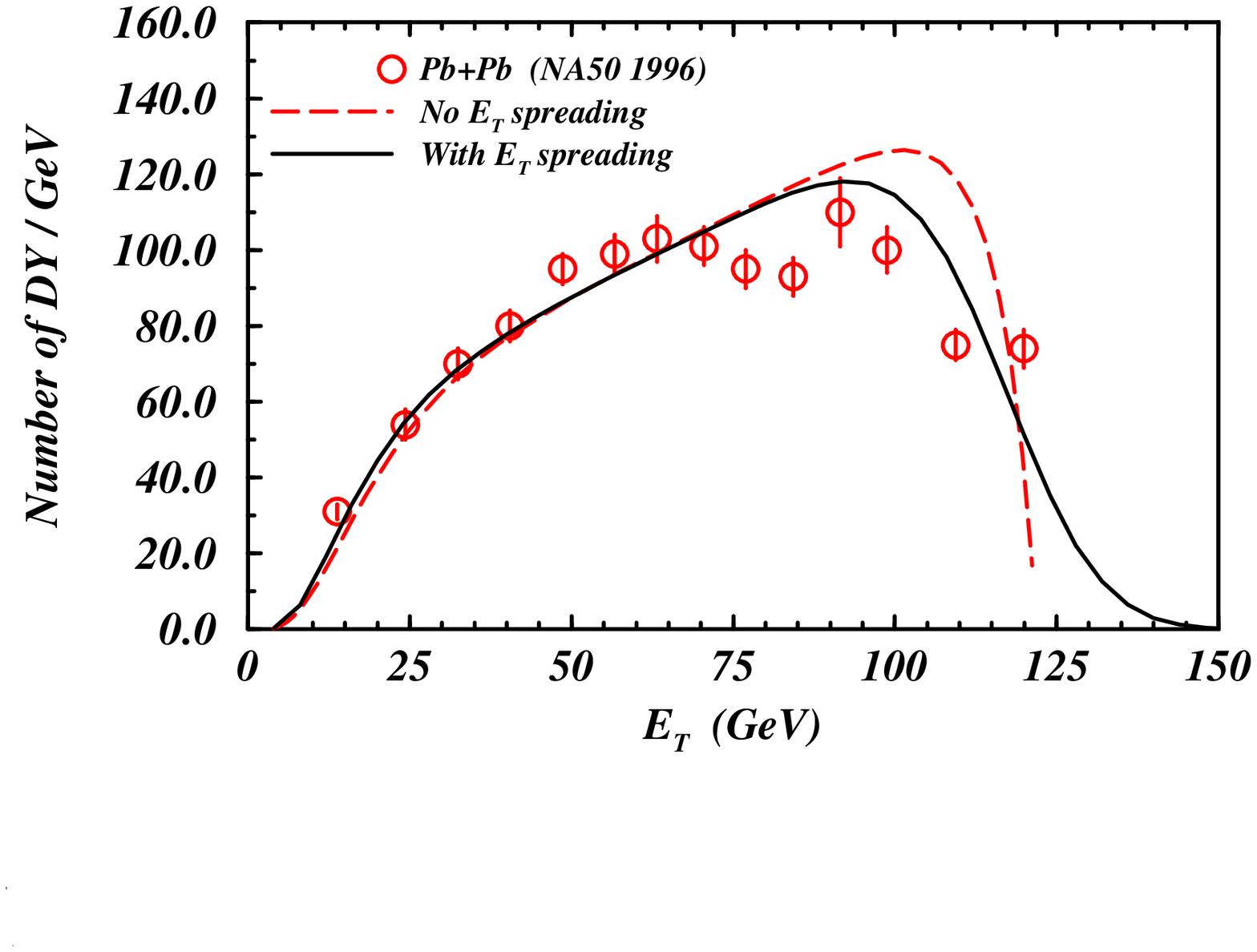}
\vskip -0.2cm
\null\hskip 0.7cm
\begin{minipage}[t]{10cm}
\noindent
\bf Fig. 1. { \rm { The Drell-Yan distribution as a function of the
transverse energy in Pb-Pb collisions at 158A GeV.  The data points
are preliminary data from NA50 \cite{Ram97,Gon98fn}.}}
\end{minipage}
\vskip 4truemm
\noindent 

One notes that for a given value of $b$, the $E_T$ value cannot be a
sharp distribution.  One can consider the righthand side of Eq.\
(\ref{eq:mean}) as defining the mean value ${\bar E}_T(b)$, and include
the spreading of $E_T$ about this mean value by introducing a
distribution function $D(b,E_T)$, 
\begin{eqnarray}
\label{eq:dis}
D(b,E_T)  = {1 \over \sqrt{2\pi} \sigma(b)} 
e^{- { [E_T-{\bar E}_T(b)]^2 / 2[\sigma(b)]^2}}.
\end{eqnarray}
Then, with the inclusion of the spreading of $E_T$, 
${d\sigma_{DY}^{AB} /dE_T}$ is given by 
\begin{eqnarray}
\label{eq:dsdetp}
{d \sigma_{DY}^{AB} \over dE_T} = A B \sigma_{DY}^{pp}
\int d\bbox{b}~ T_{AB}(\bbox{b})~ D(b,E_T) .
\end{eqnarray}
From the results of Ramello et al.\cite{Ram97,Gon98fn}, $\sigma(b)$ in
 Eq.\ ({\ref{eq:dis}) can be parameterized as
\begin{eqnarray}
\sigma(b) = {\sigma_0 \over  1 + \exp[ (b-b_\sigma)/a_\sigma]},
\end{eqnarray}
where $\sigma_0=$ 11 GeV, $b_\sigma=10$ fm, and $a_\sigma=2$ fm.  The
Drell-Yan distribution with the inclusion of $E_T$ spreading is shown
as the solid curve in Fig.\ 1, which agrees well with the preliminary
NA50 data \cite{Rom98,Gon98fn}.  The Drell-Yan distribution provides a
relation between $E_T$ and $b$ which is needed to analyze the
transverse-energy dependence of $J/\psi$ and $\psi'$ production.

\vspace*{-0.2cm}
\section{Model of $J/\psi$ and $\psi'$ Absorption}
\vspace*{-0.0cm}

We can use the $J/\psi$ absorption model presented previously in
[~\cite{Won96a}] to study $J/\psi$ and $\psi'$ absorption.  In this
model, a produced $J/\psi$ (or its precursor) meets the projectile and
target nucleon at high energies and is absorbed with a cross section
$\sigma_{\rm abs}(J/\psi\!-\!\!N)$.  It also collides with produced
hadrons (comovers) at low relative energies and is absorbed with a
cross section $\sigma_{\rm abs}(J/\psi\!-\!h)$.  In a nucleus-nucleus
collision, each nucleon-nucleon collision is a possible source of
$J/\psi$ and $\psi'$ precursors.  It is also the source of a fireball
of produced hadrons which can absorb $J/\psi$ and $\psi'$ precursors
produced by other nucleon-nucleon collisions.  One follows the
space-time trajectories of precursors, baryons, and the centers of the
fireballs of produced hadrons.  Absorption occurs when the space-time
trajectories of the precursors cross those of other particles.  Using
a row-on-row picture in the center-of-mass system and assuming
straight-line space-time trajectories, we obtain the differential
cross section for $J/\psi$ production in an $AB$ collision as
\cite{Won96a}
\begin{eqnarray}
\label{eq:fin}
{ d \sigma_{{}_{J/\psi}}^{{}^{AB}} ({\bf b}) \over \sigma_{{}_{J/\psi
}}^{{}^{NN}}~d{\bf b} } &=& \!\! \int \!\! { d{ {\bf b}}_{{}_A} \over
\sigma_{\rm abs}^2(J/\psi\!- \!\!N) } \biggl \{ 1 -\biggl [ 1-
T_{{}_A}({\bf b}_{{}_A}) \sigma_{\rm abs}(J/\psi\!-\!\!N) \biggr ] ^A \biggr
\} \nonumber\\
& &~~\times \biggl \{ 1 -\biggl [ 1- T_{{}_B}({\bf b}-{\bf b}_{{}_A})
\sigma_{\rm abs}(J/\psi\!-\!\!N)\biggr ] ^B \biggr \} F({\bf b}_A, {\bf b})\,,
\end{eqnarray}
where $F({\bf b}_A, {\bf b})$ is the survival probability due to soft
particle collisions.  To calculate $F({\bf b}_A, {\bf b})$, we sample
the target tranverse coordinate ${\bf b}_A$ for a fixed impact
parameter ${\bf b}$ in a row with the nucleon-nucleon inelastic cross
section $\sigma_{in}$.  In this row, $BT_B({\bf b}-{\bf
b}_A)\sigma_{in}$ projectile nucleons will collide with $AT_A({\bf
b}_A)\sigma_{in}$ target nucleons.  We construct the space-time
trajectories of these nucleons to locate the position of their
nucleon-nucleon collisions.  These collisions are the sources of
$J/\psi$ and $\psi'$ precursors and the origins of the fireballs of
produced particles.  For each precursor source from the collision $j$
and each absorbing fireball from the collision $i$ at the same spatial
location, we determine the time $t_{ij}^h$ when the precursor source
coexists with the absorbers in the state of produced hadrons.  The
survival probability due to this combination of precursor source and
absorber is then $\exp \{-k_{\psi h} t_{ij}^h \}$, where the rate
constant $k_{\psi h}$ is
\begin{eqnarray}
\label{eq:v1}
k_{\psi h}= \rho_h^{NN} v_h \sigma_{\rm abs}(J/\psi\!-\!h),
\end{eqnarray}
$v_h$ is the average $(J/\psi)$-$h$ relative velocity, $\rho_h^{NN}$
is the average produced hadron number density per $NN$ collision,
\begin{eqnarray}
\rho_h^{NN}= {dN_h^{NN} \over dy } { 1 \over 
\sigma_{in} (d/\gamma) },
\end{eqnarray}
$dN_h^{NN}/dy$ is the particle multiplicity per unit of rapidity at
$y_{{}_{CM}}=0$ for an $NN$ collision, $d \sim 2.46$ fm is the
internucleon spacing, and $\gamma=\sqrt{s}/2m_{\rm nucleon}$ is the
Lorentz contraction factor.  We can use the parametrization of Eq.\
(2.4) in Reference [~\cite{Won89}] to obtain ${dN_h^{NN} / dy }$ as a
function of the collision energy.

When we include all possible precursor sources and absorbers, $F({\bf
b}_A,{\bf b})$ becomes
\vskip -1.cm
\begin{eqnarray}
\label{eq:fb}
F({\bf b}_A, {\bf b}) \!= \!\sum_{n=1}^{N_<}  \!{ a(n)\over {N}_> {N}_<} 
\!\sum_{j=1}^n
\!\exp\{- \theta \!\!\!\! \sum_{i=1, i\ne j}^n 
\!\!\!k_{ \psi h} t_{ij}^h  \} \,,
\end{eqnarray}
where $N_>({\bf b}_A)$ and $N_<({\bf b}_A)$ are the greater and the
lesser of the (rounded-off) nucleon numbers $AT_A({\bf
b}_A)\sigma_{in}$ and $BT_B({\bf b}-{\bf b}_A)\sigma_{in}$, $a(n)=2
{\rm~~for~~} n=1,2,...,N_<-1 $, and $ a(N_<)=N_>-N_<+1$.  The function
$ \theta$ is zero if $A=1$ or $B=1$ and is 1 otherwise.  

We assume that the Bjorken-type longitudinal expansion occurs after
the last nucleon-nucleon collision takes place at that spatial point.
The survival probability $F$ can be determined from $k_{\psi h}$ and
\begin{eqnarray}
t_{ij}^h = t_n+t_f - {\rm Max}(t_i+t_h, t_j+t_{c\bar c}).
\end{eqnarray}
Here we shall set the $c\bar c$ production time $t_{c\bar c}$ equal to 0.06
fm/c, the hadronization time $t_h$ equal to 1.2 fm/c, and $t_f$ is
related to the freezeout time $t_{\rm freezeout}$ by
\begin{eqnarray}
t_f=t_h+t_h \ln( t_{\rm freezeout}/t_h)
 = t_h+t_h \ln(\rho_h^{NN}/ \rho_{\rm freezeout}^{NN}).
\end{eqnarray}
We use a freezeout density of $\rho_{\rm freezeout}^{NN}=0.5$
hadrons/fm${}^3$, corresponding to a produced hadron freezeout
separation of 1.3 fm.  The $\psi'$ production cross section can be
obtained from the above equations by changing $J/\psi$ into $\psi'$.

We have used this absorption model to study the experimental $J/\psi$
and $\psi'$ data in $p$-A and nucleus-nucleus collisions
\cite{Won96qm,Won97}.  We found that $J/\psi$ absorption by produced
hadrons, as revealed by O-Cu, O-U, and S-U collisions, is small, and
the experimental Pb-Pb yield is much smaller than the extrapolated
results if one assumes $J/\psi$ absorption by nucleons and produced
hadrons only.

The deviation of the $J/\psi$ data in Pb-Pb collisions from the
conventional theoretical extrapolations of $p$-A, O-A, and S-U
collisions suggests that there is a transition to a new phase of
strong absorption which sets in when the local energy density exceeds
a certain threshold.  We have extended the absorption model to
describe this transition in terms of the critical number of $NN$
collisions at a spatial point \cite{Won96qm,Won97}.  We can
reformulate the model here in terms of the critical energy density.
Evaluated in the nucleon-nucleon center-of-mass system, the energy
density at the spatial point, which has had $\nu$ prior
nucleon-nucleon collisions, is approximately
\vskip -0.5cm
\begin{eqnarray}
\epsilon= \nu \rho_h^{NN} m_t 
\end{eqnarray}
\vskip -0.1cm\noindent where $m_t$ is the average transverse mass of a
produced hadron.

\null\vskip 3.2cm 
\epsfxsize=170pt
\includegraphics{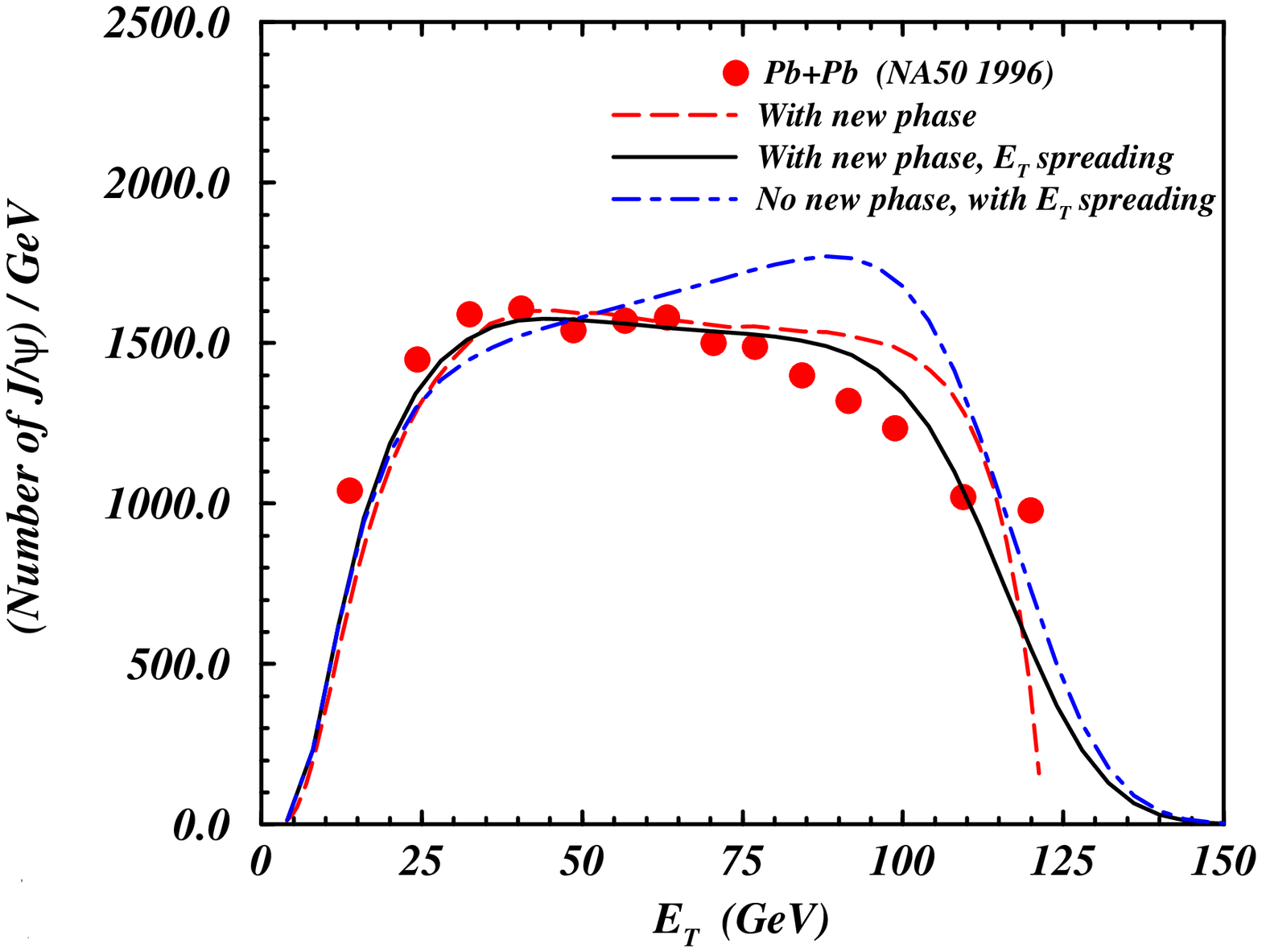}
\vspace*{2.2cm}
\null\hskip 0.7cm
\begin{minipage}[t]{10cm}
\noindent
\bf Fig. 2. { \rm { The $J/\psi$ distribution as a function of the
transverse energy in Pb-Pb collisions at 158A GeV. The data points are
preliminary data from NA50 \cite{Ram97,Gon98fn}.}}
\end{minipage}
\vskip 4truemm
\noindent 

We postulate that soft particles make a transition
to a new phase with stronger $J/\psi$ absorption at a spatial point if
the energy density at that point exceeds $\epsilon_{\rm crit}$.  When
this happens, the matter in the new phase absorbs $J/\psi$ with a
greater strength.  To keep track of the same matter density, we still
use the same number of particles in the new phase as in the hadron
matter, but the particles in the new phase absorb $J/\psi$ with an
effective cross section $\sigma_{\rm abs}(J/\psi$-$x)$.  The quantity
$ k_{\psi h} t_{ij}^h$ in Eq.\ (\ref{eq:fb}) becomes $ k_{\psi h}
t_{ij}^h + k_{\psi x} t_{ij}^x$, where the new rate constant $k_{\psi
x}$ is
\begin{eqnarray}
\label{eq:v2}
k_{\psi x}= \rho_h^{NN} v_x \sigma_{\rm abs}(J/\psi\!-\!x),
\end{eqnarray}
and the quantity $t_{ij}^x$ is the time for a $J/\psi$ produced in
collision $j$ to coexist at the same spatial location with the
absorbing soft particles produced in collision $i$ in the form of the
new phase.  The duration when the matter is in the new phase can be
inferred by assuming that the matter undergoes Bjorken-type
longitudinal expansion after all the nucleon-nucleon collisions have
occurred at that point.  Because the soft particles are relativistic, we
take all relevant velocities $v$ in Eqs.\ (\ref{eq:v1}) and
(\ref{eq:v2}) to be 1.

We obtain good agreement with Pb-Pb data using the following
parameters: $\sigma_{\rm abs} (\psi\!-\!N)=\sigma_{\rm abs}
(\psi'\!-\!N)=6.36 {\rm ~~mb}$, $\epsilon_{\rm crit}= 4.2 {\rm~~
GeV/fm}^3$, $\sigma_{\rm abs}(J/\psi\!-\!\!h)=0.15$ mb, $ \sigma_{\rm
abs}(J/\psi\!-\!\!x)=3.7$ mb, and $\sigma_{\rm abs}(\psi'\!-\!\!h)=$
11 mb. Note that $\sigma_{\rm abs}(J/\psi\!-\!\!x) >> \sigma_{\rm
abs}(J/\psi\!-\!\!h)$.  We show in Fig.\ 2 the $J/\psi$ distribution
in Pb-Pb collisions at 158A GeV.  The solid curve is the result when
one assumes a new phase and $E_T$ spreading, the dashed curve is the
result with the new phase but without $E_T$ spreading.  The dashed-dot
curve is the result without the new phase, but the $E_T$ spreading is
included.  There is good agreement of the theoretical curve with the
preliminary NA50 data \cite{Rom98} when one assumes a new phase of
strong $J/\psi$ absorption and takes into account the $E_T$ spreading.

\null\vskip 7.1cm \epsfxsize=250pt
\includegraphics{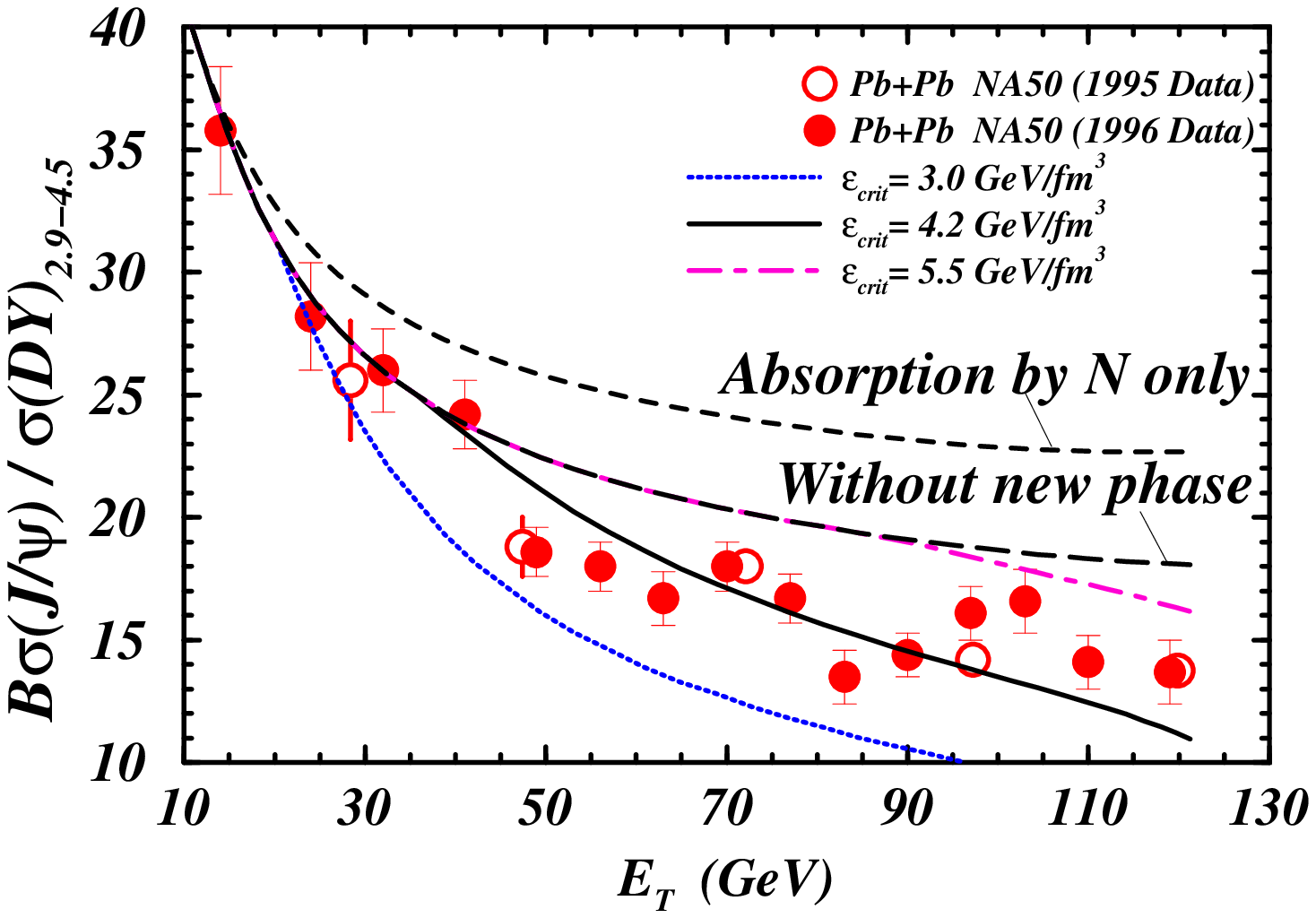}
\vspace*{-1.1cm}
\null\hskip 0.8cm
\begin{minipage}[t]{10cm}
\noindent
\bf Fig. 3. { \rm { The ratio of ${\cal B}\sigma(J/\psi)/\sigma(DY)$
for Pb-Pb collisions at 158A GeV as a function of $E_T$. The data
points are preliminary data from NA50 \cite{Rom98}.}}
\end{minipage}
\vskip 4truemm
\noindent 

In Fig.\ 3 we show ${\cal B}\sigma^{AB}(J/\psi)/\sigma^{AB}(DY)$ for
Pb-Pb collisions at 158A GeV.  The short-dashed curve gives the
results when there is absorption by nucleons only, while the
long-dashed curve gives the results when one assumes absorption by
nucleons and produced hadrons.  Depending on the critical energy
density, the ratio ${\cal B}\sigma^{AB}(J/\psi)/\sigma^{AB}(DY)$
deviates from the long-dashed curve (for no new phase) at different
transverse energies.  For $\epsilon_{\rm crit}=4.2$ GeV/fm$^3$, the
break occurs at $E_T \sim 40 $ GeV which is close to the break
observed experimentally\cite{Rom98}.  For $\epsilon_{\rm crit}=3.0$
GeV/fm$^3$, it occurs at $E_T\sim 20$ GeV, and for $\epsilon_{\rm
crit}=5.5$ GeV/fm$^3$, it occurs at $E_T\sim 90$ GeV.  However, the
discontinuity in the theoretical results is much smoother than the
discontinuity in the preliminary NA50 data \cite{Rom98} shown in Fig.\
3.

\vspace*{-0.3cm}
\section{Dependence of $J/\psi$ Survival Probability on Collision Energy}
\vspace*{-0.1cm}

The above formulation allows one to calculate the $J/\psi$ survival
probabilities, represented by $S(b)=[{\cal B}\sigma^{AB}(J/\psi)/
\sigma^{AB}(DY)] /[{{\cal B} \sigma^{pp}(J/\psi) }/{
\sigma^{pp}(DY)}]$, as a function of the impact parameter for
different collision energies.  The results are shown in Fig.\ 4.  The
short-dashed curve is the survival probability when $J/\psi$ is
absorbed by nucleons only.  The dash-dot and the solid curves are the
results when there is a new phase.   The dashed curves are the
survival probabilities when there is no new phase.  The numbers
label the collision energy, $\sqrt{s}$.

\null\vskip 3.4cm \epsfxsize=250pt
\includegraphics{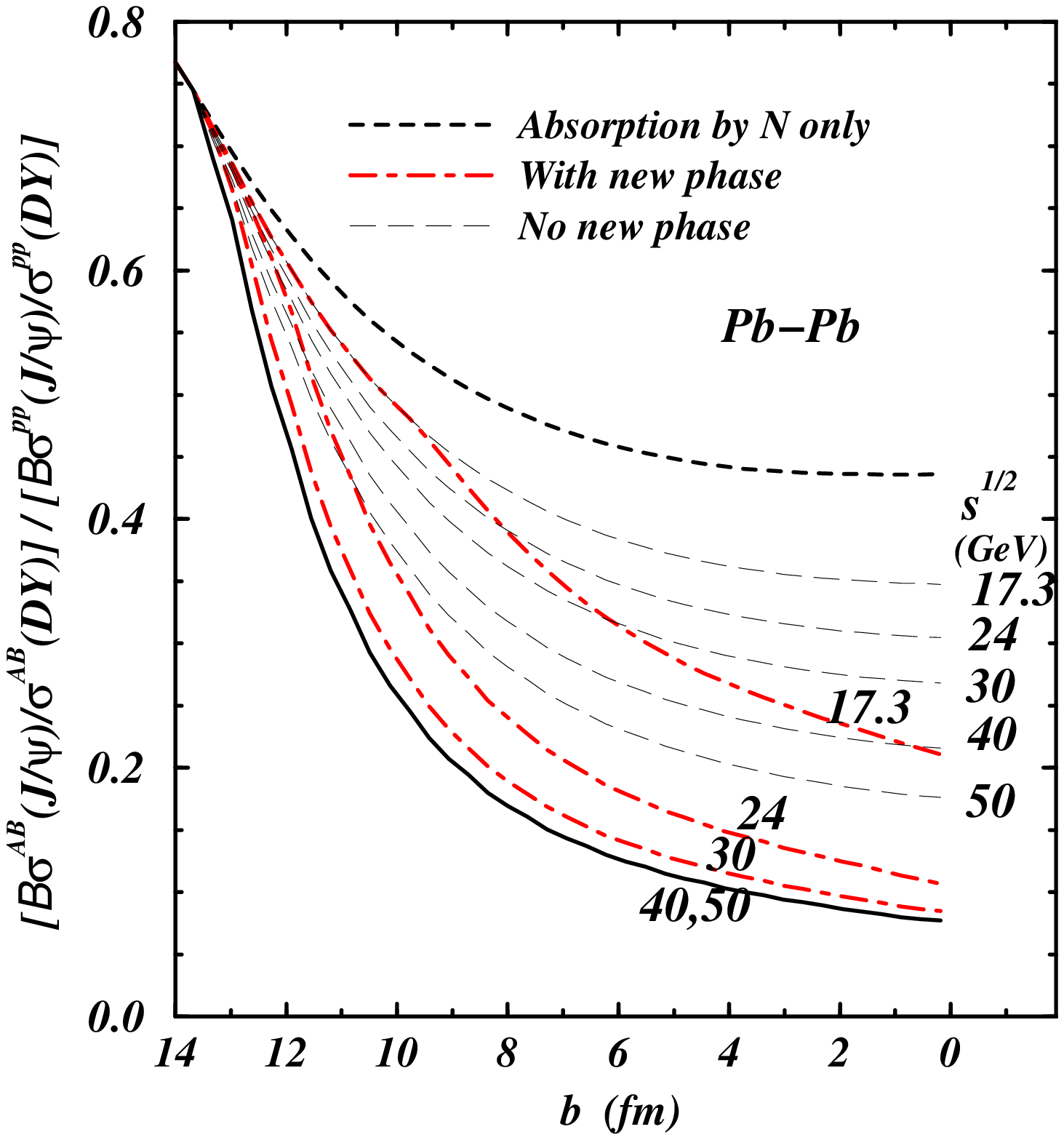}
\vskip 4.8cm
\null\hskip 0.5cm
\begin{minipage}[t]{10.5cm}
\noindent
\bf Fig. 4. { \rm { The ratio $[{\cal B}\sigma^{AB}(J/\psi)/
\sigma^{AB}(DY)] /[{{\cal B} \sigma^{pp}(J/\psi) }/{
\sigma^{pp}(DY)}]$ for Pb-Pb collisions at different collision
energies as a function of the impact parameter.  }}
\end{minipage}
\vskip 4truemm
\noindent 

Figure 4 shows that if a new phase of strong $J/\psi$ absorption is
assumed, the $J/\psi$ survival probability reaches the ``lowest
survival'' (LS) limit (the solid curve) when the nucleon-nucleon
center-of-mass energy $\sqrt{s}$ reaches about 35 GeV.  This is the
limit at which the energy density of the produced matter is so large
that no $J/\psi$ particles, except those produced at the peripheral
region of the collision, can survive the nucleus-nucleus collision.
If there is no phase transition, the lowest survival limit is reached
at about $\sqrt{s} \sim 70$ GeV.

The existence of this lowest $J/\psi$ survival limit shows up as a
universal curve when one plots the $J/\psi$ survival probability as a
function of $b$ for different energies.  This can be obtained by
transforming $S(E_T)$ to $S(b)$, utilizing the method of Section 2 to
make a correspondence between $E_T$ and $b$ by means of the Drell-Yan
distribution.  Upon verification of the existence of the lowest
survival limit, the energy at which this LS limit is reached then
gives a good indication as to whether the new phase of strong absorption
occurs or not.

The above results have an important implication for the RHIC collider.
At the energy $\sqrt{s}=200$ GeV, one expects that the $J/\psi$
survival probability, as measured by $S(b)=[{\cal
B}\sigma^{AB}(J/\psi)/ \sigma^{AB}(DY)] /[{{\cal B}
\sigma^{pp}(J/\psi) }/{ \sigma^{pp}(DY)}]$, will be in the lowest
survival limit.  However, such a limit can also be reached by the
absorption of $J/\psi$ by collisions with produced hadrons alone,
as one can see from Fig.\ 4.  It will be difficult to distinguish the
two cases at such a high energy.  It is therefore important to perform
experiments at lower energies where such a distinction can be
observed.

\section{Conclusions}

Using the relation between $E_T$ and $b$ obtained from the Drell-Yan
distribution, we have analyzed the transverse-energy dependence of
$J/\psi$ production.  We found that the anomalous suppression of
$J/\psi$ in Pb-Pb collisions can be explained by a model in which a
new phase of strong absorption sets in when the local energy density
exceeds 4.2 GeV/fm${}^3$.  The critical energy density $\epsilon_{\rm
crit}$ which corresponds to the onset of the new phase is close to the
quark-gluon plasma energy density, $\epsilon_c \sim 4.17$ GeV/fm$^3$,
calculated from the lattice gauge theory result \cite{Blu95} of
$\epsilon_c/T_c^4 \sim 20$ with $T_c \sim 0.2$ GeV.  Therefore, it is
interesting to speculate whether the new phase of strong absorption
may be the quark-gluon plasma.  However, other signatures are needed
to corroborate whether the quark-gluon plasma has been produced.

Our formulation of the absorption model in terms of produced hadron
densities allows us to calculate $J/\psi$ survival probabilities at
higher energies.  We found interesting new results that the $J/\psi$
survival probability reaches the lowest survival limit when the
collision energy $\sqrt{s}$ reaches about 35 GeV.  At the energy of
$\sqrt{s}=200$ GeV, it will become difficult to distinguish absorption
by the new phase or absorption by produced soft particles, as both
effects will lead to the lowest survival limit.

\vspace*{-0.3cm}
\section*{Acknowledgments}
\vspace*{-0.1cm}

This research was supported by the Division of Nuclear Physics,
U.S. D.O.E.  under Contract DE-AC05-96OR22464 managed by Lockheed
Martin Energy Research Corp.



\vspace*{-0.3cm}
\section*{References}
\vspace*{-0.1cm}


\begin{thebibliography}{99}

\bibitem{Mat86} T. Matsui and H. Satz, Phys. Lett. B178 (1986) 416.

\bibitem{Gon96} M. Gonin, NA50 Collaboration, Nucl. Phys. 
A610 (1996) 404c.

\bibitem{Lou96} C. Louren\c co, NA50 Collaboration, Nucl. Phys. 
A610 (1996) 552c.

\bibitem{Ram97} L. Ramello $et~al.$, NA50 Collaboration, in
Proceedings of Quark Matter '97 Conference, Tsukuba, December 1-5,
1997.

\bibitem{Rom98} A. Romana $et~al.$, NA50 Collaboration, in Proceedings
of the XXXIII Recontres de Moriond, Les Arcs, France, 21-28 March,
1998.

\bibitem{Won96qm} C. Y. Wong, Nucl. Phys.  A610 (1996) 434c.

\bibitem{Kha96qm} D. Kharzeev, Nucl. Phys.  A610 (1996) 418c

\bibitem{Bla96qm} J.-P. Blazoit and J.-Y. Ollitrault, 
Nucl. Phys.  A610 (1996) 452c.

\bibitem{Gav96qm} S. Gavin and R. Vogt,
Nucl. Phys.  A610 (1996) 442c.

\bibitem{Won97} C. Y. Wong, Phys. Rev. C55 (1997) 2621;
C. Y. Wong, Nucl. Phys. A630, 487 (1998); C. Y. Wong, hep-ph/9712332.

\bibitem{Cap96} A. Capella, A. Kaidalov, A. K. Akil, and C. Gerschel,
Phys. Lett. B393 (1997) 431; N. Armesto, A. Capelle, and
E. G. Ferreiro, hep-ph/9807258.

\bibitem{Cas96} W. Cassing and C. M. Ko, Phys. Lett. B396 (1997) 39. 

\bibitem{Kha96a} D. Kharzeev, C. Louren\c co, M. Nardi, and H. Satz,
Z.Phys. C74, 307 1997; D. Kharzeev, nucl-th/9802037.

\bibitem{Arm96}
N. Armesto, M. A. Braun, E. G. Ferreiro, and C. Pajares,
Phys. Rev. Lett. 77 (1996) 3736.

\bibitem{Won96a} 
C. Y. Wong, Phy. Rev. Lett. 76  (1996) 196.

\bibitem{Ald91}
D. M. Alde $et~al.$, E772 Collaboration, Phys. Rev. Lett. 66,
133 (1991). 

\bibitem{Won94} C. Y. Wong, {\it Introduction to High-Energy Heavy-Ion
    Collisions}, World Scientific Publishing Company, 1994.

\bibitem{Gon98fn} The transverse energies in Ref. [~\cite{Ram97}] need
to be rescaled by a factor of 0.81 to give the transverse energies in
Ref.\ [ \cite{Rom98}] (M. Gonin, Charmonium Workshop, Institute of
Nuclear Theory, Seattle, May 18-22, 1998).  Therefore, we multiply the
transverse energies of Ref. [ \cite{Ram97}] by 0.81 in our analysis.

\bibitem{Won89} C. Y. Wong and Z. D. Lu, Phys. Rev. D39, 2606 (1989).

\bibitem{Blu95} T. Blum $et~al.$, Phys. Rev. D51 (1995) 5153.

\end{thebibliography}
\end{document}